\documentclass[prl,twocolumn,preprintnumbers,superscriptaddress,aps,amsmath,amssymb,showpacs,nofootinbib]{revtex4-2}

\usepackage{graphicx} 
\usepackage{hyperref}
\usepackage{braket}
\usepackage{slashed}
\usepackage{tikz-feynman}
\usepackage{comment}

\newcommand{\tr}{\mathrm{tr}}

\usepackage[normalem]{ulem}

\begin{document}
\preprint{PITT-PACC-2607}
\preprint{CPTNP-2026-018}

\title{Quantum Information of Photon Pairs at Lepton Colliders}

\author{Kun Cheng}
\email{kun.cheng@pitt.edu}
\affiliation{PITT PACC, Department of Physics and Astronomy,\\ University of Pittsburgh, 3941 O’Hara St., Pittsburgh, PA 15260, USA}

\author{Tao Han}
\email{than@pitt.edu}
\affiliation{PITT PACC, Department of Physics and Astronomy,\\ University of Pittsburgh, 3941 O’Hara St., Pittsburgh, PA 15260, USA}

\author{Guanghui Li}
\email{ghli@ihep.ac.cn}
\affiliation{Institute of High Energy Physics, Chinese Academy of Sciences, Beijing 100049, China}
\affiliation{School of Physical Sciences, University of Chinese Academy of Sciences, Beijing 100049, China}

\author{Bin Yan}
\email{yanbin@ihep.ac.cn (corresponding author)}
\affiliation{Institute of High Energy Physics, Chinese Academy of Sciences, Beijing 100049, China}
\affiliation{Center for High Energy Physics, Peking University, Beijing 100871, China}

\begin{abstract}
Photon pairs have provided an ideal laboratory for exploring entanglement and Bell inequality violation in low-energy experiments. Extending such studies to high-energy colliders is of great interest but has yet to be explored. Exploiting the photon conversion process for nearly on-shell photons, we formulate a factorization framework and an effective two-qubit description, which enable access to quantum information encoded in photon pairs.
Using the existing Belle data set, we estimate that a $7.4\sigma$ violation of the Bell inequality could be achieved.  The same framework can also probe quantum discord and nonstabilizerness, which could be measured with precisions of 5.6\% and 1.6\%, respectively. All the reconstructed results from photon conversion in the two-qubit framework are found to be consistent with the kinematic approach of real photons, and the formalism can apply to other spin-1 systems in an appropriate two-qubit limit.
\end{abstract}

\maketitle

\noindent
\emph{Introduction ---}
Photons have provided an ideal platform for studying the most characteristic quantum phenomena, enabling early measurements of spin correlations between two particles~\cite{Wu:1950zz}, pioneering tests of Bell inequalities~\cite{Freedman:1972zza,kasday1975angular,Aspect:1982fx}, and establishing the violation of Bell inequalities and ushering quantum information science \cite{Weihs_1998}.

Recent progress in studying quantum correlation in high-energy collisions has opened a new avenue in exploring quantum information science with a plethora of data in high-energy experiments 
\cite{Afik:2020onf,Fabbrichesi:2021npl,Severi:2021cnj,Afik:2022kwm,Afik:2022dgh,Han:2023fci,Aguilar_Saavedra_2022,Dong_2024,Cheng:2023qmz,Cheng:2024btk,Han:2024ugl,Afik:2026pxv,Cheng:2024btk,Altakach:2022ywa,Ehataht:2023zzt,Ma:2023yvd,Fabbrichesi:2024wcd,Han:2025ewp,Morales:2024jhj,Zhang:2025mmm,Ai:2025wnt,Barr:2021zcp,Barr:2022wyq,Ashby-Pickering:2022umy,Aguilar-Saavedra:2022wam,Fabbrichesi:2023cev,Fabbri:2023ncz,Bi:2023uop,Morales:2023gow,Bernal:2024xhm,Grossi:2024jae,Goncalves:2025qem,Goncalves:2025xer,Aguilar-Saavedra:2022mpg,Aguilar-Saavedra:2025byk,Ding:2025mzj,ATLAS:2026hye,CMS:2025anw,Du:2024sly,Afik:2025grr,Chen:2013epa,Pei:2025yvr,Pei:2025ito,Fucilla:2025kit,Cheng:2025cuv,Lin:2025eci,Cheng:2025zcf,Guo:2026yhz,Fang:2026ddi,Qi:2025onf,Cheng:2025zaw,Hatta:2025obw,Fucilla:2025kit,Ruzi:2025jql,Zhang:2025mmm,Cao:2025qua,Feng:2025ryr,Fucilla:2026mkg,Zhang:2026nwm,Liu:2026dzv,Wang:2026nls,Goncalves:2026njf,Du:2026qsy,Yang:2026uwu,Li:2026bkf,Pei:2026rlh,Pei:2026wfu}, signified by the observation of spin entanglement of top quark pair~\cite{ATLAS:2023fsd,CMSCollaboration_2024} and $H\to ZZ^*\to\ell^+\ell^-\ell^+\ell^-$~\cite{ATLAS:2026hye,CMS:2025anw} at the Large Hadron Collider (LHC).
It is understood that, given the particle identification capabilities of the current detector technologies, high-energy experiments do not offer Alice-Bob's setting originally proposed by John Bell \cite{PhysicsPhysiqueFizika.1.195} to test realism or to generally exclude interpretation by local hidden variable theories~\cite{Abel:1992kz,Li:2024luk,Bechtle:2025ugc,Abel:2025skj,Low:2025aqq,Aguilar-Saavedra:2026rsx}. Instead, the focus has been on understanding quantum information within the framework of quantum field theory and exploring new interactions beyond the Standard Model~\cite{Afik:2025ejh}.

There are two complementary approaches to establishing the spin-entanglement with high-energy data. One is to rely on the production of a quantum state without requiring the subsequent decays, dubbed the ``kinematic approach'' \cite{Cheng:2024rxi}. This is the original method for extracting spins for light quarks \cite{Hanson:1975fe} and gluons \cite{Ellis:1978wp,TASSO:1980lqw}. Without the capability of a polarizer, the spin information of the photons and electrons observed in an electromagnetic calorimeter or a tracker would only be inferred from kinematic distributions. 
Another more informative approach is to follow the decays after the production of the quantum state, including $t\bar t$ \cite{Afik:2020onf,Fabbrichesi:2021npl,Severi:2021cnj,Afik:2022kwm,Afik:2022dgh,Han:2023fci,Aguilar_Saavedra_2022,Dong_2024,Cheng:2023qmz,Cheng:2024btk,Han:2024ugl,Afik:2026pxv}, $\tau^+\tau^-$ \cite{Altakach:2022ywa,Ehataht:2023zzt,Ma:2023yvd,Fabbrichesi:2024wcd,Han:2025ewp,Morales:2024jhj,Zhang:2025mmm,Ai:2025wnt}, and vector bosons~\cite{Barr:2021zcp,Barr:2022wyq,Ashby-Pickering:2022umy,Aguilar-Saavedra:2022wam,Fabbrichesi:2023cev,Fabbri:2023ncz,Bi:2023uop,Morales:2023gow,Bernal:2024xhm,Grossi:2024jae,Goncalves:2025qem,Goncalves:2025xer,Aguilar-Saavedra:2022mpg,Aguilar-Saavedra:2025byk,Ding:2025mzj,ATLAS:2026hye,CMS:2025anw}.
Along this approach, we consider a quantum ensemble of correlated photon pairs as a bi-qubit system and propose to obtain quantum information of photons by exploiting photon conversion into lepton pairs $\gamma^*\to \ell^+\ell^-$.
This spin correlation has proven to be important in determining fundamental quantum numbers such as the CP property of $\pi^0$~\cite{Plano:1959zz,Samios:1962zza}.

For a spin-1 vector boson, the massive particle (or an off-shell photon) has three polarization states, thus effectively forming a qutrit. However, in certain circumstances, a vector boson can be simplified as a qubit while preserving all quantum information. We demonstrate this by exploiting a nearly on-shell photon conversion, which can be factorized from the hard production dynamics, and the two physical polarization states form an effective qubit. 

In this Letter, we develop a factorization and quantum tomography framework for photon conversion processes, which reconstructs the linear polarizations of the two-photon polarization density matrix of two qubits in the nearly on-shell limit. This thereby enables and significantly simplifies collider measurements of quantum information of photons.
As an application, we study photon pair production in Belle  \cite{Belle:2000cnh} and find that a violation of the Clauser–Horne–Shimony–Holt (CHSH) inequality~\cite{Clauser:1969ny} could be established at the level of $7.4\sigma$ using the existing data set, with the improved sensitivity expected in the Belle II experiment due to its larger data sample \cite{Belle-II:2010dht}.
For configurations with negatively correlated photon helicities in the on-shell limit, the same framework can also be applied to other quantum information observables. In particular, quantum discord could be measured to a precision of 5.6\%, while nonstabilizerness could be measured to 1.6\%. 
We demonstrate the consistency between reconstructed results and the kinematic approach, which, for the first time, shows the validity of the qubit treatment of virtual photons from the quantum information aspect.

\vskip 0.3mm
\noindent
\emph{Spin state for nearly on-shell photon pair ---}
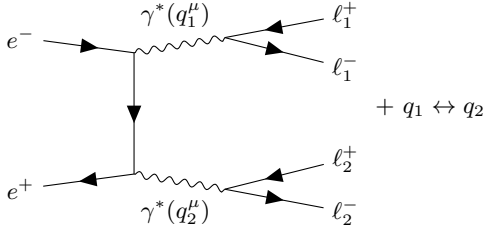
\begin{figure}
    \centering
        \begin{tikzpicture}[baseline=(center.base)]
        \begin{feynman}
            \vertex (i1) at (-1.5cm,1.0cm){$e^-$};
            \vertex (i2) at (-1.5cm,-1.0cm){$e^+$};
            \vertex (v1) at (0cm,0.8cm);
            \vertex (v2) at (0cm,-0.8cm);
            \vertex (o1) at (1.2cm,1.0cm);
            \vertex (o2) at (1.2cm,-1.0cm);
            \vertex (t1) at (2.8cm,1.3cm){$\ell_1^+$};
            \vertex (t2) at (2.8cm,0.6cm){$\ell_1^-$};
            \vertex (t3) at (2.8cm,-0.6cm){$\ell_2^+$};
            \vertex (t4) at (2.8cm,-1.3cm){$\ell_2^-$};
            \vertex (q1) at (0.55cm,1.3cm){$\gamma^*(q_1^\mu)$};
            \vertex (q2) at (0.55cm,-1.3cm){$\gamma^*(q_2^\mu)$};
            \vertex (center) at (0,0){};
            \diagram{(i1)--[fermion](v1)--[fermion](v2)--[fermion](i2),(v1)--[boson](o1),(v2)--[boson](o2),(t1)--[fermion](o1)--[fermion](t2),(t3)--[fermion](o2)--[fermion](t4)};
        \end{feynman}
    \end{tikzpicture} + $q_1 \leftrightarrow q_2$
    \caption{Feynman diagrams for photon-pair production at a lepton collider followed by virtual-photon conversions into lepton pairs.}
    \label{fig:FeynDiag}
\end{figure}
We consider a correlated two-photon system produced in 
\begin{equation}
    e^+e^-\to \gamma^*\gamma^* \to \ell_1^+ \ell_1^- \ \ell_2^+ \ell_2^-, 
\end{equation}
as shown in Fig.~\ref{fig:FeynDiag}. The two virtual photons are produced and subsequently converted into lepton pairs. The invariant mass of each photon is given by $q_i^2=m_{\ell_i^+\ell_i^-}^2$, and we define the virtuality of the photon with respect to the scattering center-of-mass (c.m.) energy $\sqrt{s}$ as $\delta_i = \sqrt{q_i^2}/\sqrt{s}$.

The propagator of a virtual photon with momentum $q$ can be decomposed as 
\begin{equation}
    g^{\mu\nu} = \frac{q^\mu q^\nu}{q^2} - \sum_{\lambda=0,\pm1} \epsilon_\lambda^\mu(q) \epsilon_\lambda^{\nu*}(q).
\end{equation}
Since the final-state lepton pair produced from each virtual photon carries identical masses, the $q^\mu q^\nu/q^2$ term does not contribute to the amplitude due to current conservation. The squared amplitude can therefore be written in a factorized form,
\begin{equation}\label{eq:qutrit}
    |\mathcal{M}|^2 = \frac{1}{q_1^4 q_2^4} \sum_{\lambda=0,\pm1} R_{\lambda_1\lambda_1^\prime, \lambda_2\lambda_2^\prime} \Gamma^{(1)}_{\lambda_1\lambda_1^\prime} \Gamma^{(2)}_{\lambda_2\lambda_2^\prime}.
\end{equation}
Here, $R_{\lambda_1\lambda_1^\prime,\lambda_2\lambda_2^\prime}$ is the unnormalized production spin density matrix of the virtual photon pair,
\begin{equation}
    R_{\lambda_1\lambda_1^\prime, \lambda_2\lambda_2^\prime} = \sum_{\sigma\bar \sigma} \mathcal{M}(e^-_\sigma e^+_{\bar \sigma} \to \gamma_{\lambda_1} \gamma_{\lambda_2}) \mathcal{M}(e^-_\sigma e^+_{\bar \sigma} \to \gamma_{\lambda_1^\prime} \gamma_{\lambda_2^\prime})^*,
\end{equation}
where $\sigma$ and $\bar\sigma$ are the helicities of the initial-state electron and positron, while $\lambda_i$ and $\lambda_i^\prime$ denote the helicities of the virtual photons. The quantity $\Gamma_{\lambda,\lambda^\prime}^{(i)}$ is the ``splitting density matrix" associated with the conversion of the $i$-th virtual photon into a lepton pair,
\begin{equation}\label{eq:Gamma0}
    \Gamma^{(i)}_{\lambda\lambda^\prime} = \sum_{s,\bar s}  \Big(\bar \ell_i\slashed{\epsilon}_\lambda(q_i) \ell_i\Big)\Big(\bar \ell_i\slashed{\epsilon}_{\lambda^\prime}(q_i) \ell_i\Big)^*,
\end{equation}
where the helicities $s$ and $\bar s$  of the final-state lepton and antilepton are summed over.
 
The spin indices of the virtual photon in Eq.~\eqref{eq:qutrit} sum over both transverse and longitudinal modes. The spin state of a virtual photon is therefore described by a qutrit state, as any massive spin-1 particle in general. However, when both photons are close to the mass shell ($\delta_i\to 0$), the longitudinal polarization vector becomes proportional to the photon momentum,
\begin{equation}
    \epsilon_0^\mu(q_i) = \frac{q_i^\mu}{\sqrt{q_i^2}} + \mathcal{O}(\delta_i),
\end{equation}
so that the longitudinal contributions to both the production density matrix $R$ and the splitting density matrix $\Gamma^{(i)}$ are suppressed. The scattering process can therefore be effectively described by the transverse components,
\begin{equation}\label{eq:M2fromRho}
    \frac{d\sigma}{d\Omega} \propto \sum_{\lambda=\pm1} R_{\lambda_1\lambda_1^\prime, \lambda_2\lambda_2^\prime} \Gamma^{(1)}_{\lambda_1\lambda_1^\prime} \Gamma^{(2)}_{\lambda_2\lambda_2^\prime} + \mathcal{O}(\delta^2).
\end{equation}

\begin{figure}
    \centering
    \includegraphics[width=0.9\linewidth]{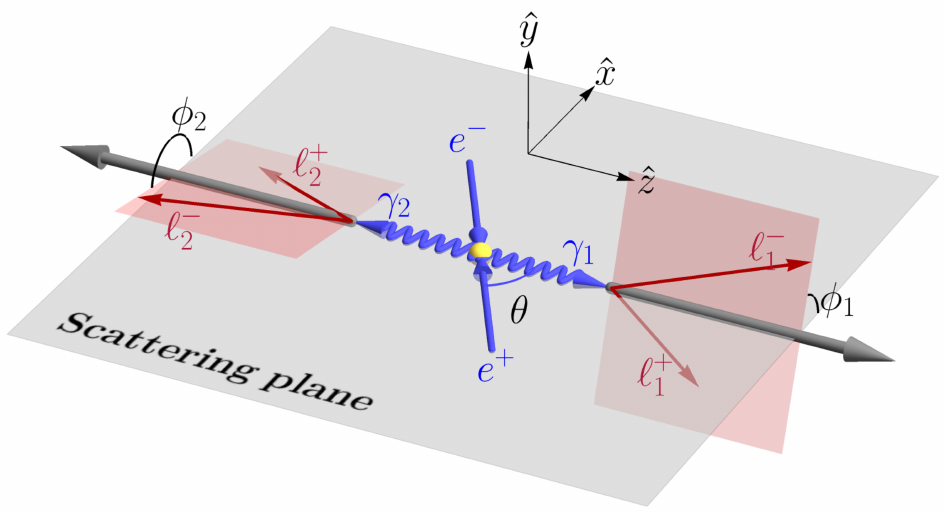}
    \caption{Leading order kinematic configuration of photon pair production with virtual-photon conversions into lepton pairs in $e^-e^+$ collisions. }
    \label{fig:illustratonPlot}
\end{figure}

Consequently, up to $\mathcal{O}(\delta^2)$ corrections, the spin state of the virtual photon pair in the process $e^-e^+\to \gamma^*\gamma^*(\to \ell_1^+ \ell_1^- \ell_2^+\ell_2^- )$ can be effectively treated as a qubit pair.
The corresponding normalized $4\times 4$ spin density matrix, $\rho=R/\tr(R)$, can be parametrized as,
\begin{equation}
    \rho= \frac{I_2\otimes I_2+ B_i^{(1)}\sigma_i\otimes I_2 + B_i^{(2)} I_2\otimes \sigma_i +C_{ij}\sigma_i\otimes\sigma_j}{4},
\end{equation}
where $I_2$ is the 2-dimensional identity matrix and $\sigma_i$ is the Pauli matrix.
Here, we choose the same spin quantization axis for both photons, with the $z$ direction defined along the momentum of $\gamma_1$ (see Fig.~\ref{fig:illustratonPlot}), and the choice of $\gamma_1$ is arbitrary due to identical particles.
Then the eigenstates of $\sigma_z$ correspond to the circular polarized photon with positive/negative helicity, while the eigenstates of $\sigma_{x/y}$ correspond to linearly polarized states.
The coefficients $B_{x/y}^{(i)}$ and $B_z^{(i)}$, respectively, quantify the degree of linear and circular polarization of the $i^{\rm th}$ photon, while $C_{ij}$ is the correlation matrix between the polarization of two photons.

Within this effective two-qubit description, the full quantum information is encoded in $B^{(1)}, B^{(2)}$ and $C_{ij}$. One of the most characteristic quantum correlations is entanglement, which can be quantitatively described by the concurrence $\mathcal{C}$, defined as $\mathcal{C}=\max\!\left\{0,\lambda_1-\lambda_2-\lambda_3-\lambda_4\right\}$ where $\lambda_i$ are the eigenvalues in the descending order of the matrix $\sqrt{\sqrt{\rho} \tilde{\rho} \sqrt{\rho}}$, with $\tilde{\rho}=(\sigma_2 \otimes \sigma_2) \rho^* (\sigma_2 \otimes \sigma_2)$~\cite{Wootters:1997id}, and $\mathcal{C}>0$ implies an entangled state.
A stronger condition than entanglement, the Bell inequality violation, can be measured with a simpler linear variable 
\begin{equation}\label{eq:CHSH}
   \mathcal{B}_\pm = C_{xx}\pm C_{yy}.
\end{equation}
The CHSH inequality states $|\mathcal{B}_\pm|<\sqrt{2}$, which can be violated by quantum mechanical predictions.
The measured value of the Bell variable Eq.~\eqref{eq:CHSH} also provides a lower bound of the entanglement between the two photons through the following relation~\cite{Verstraete:2001skx},
\begin{equation}\label{eq:CBrelation}
    \mathcal{C}\geq \sqrt{\frac{\mathcal{B}_\pm^2}{2}-1}.
\end{equation}

\vskip 0.3mm
\noindent 
\emph{Photon polarization from final state distribution ---}
With Eq.~\eqref{eq:M2fromRho}, the photon spin density matrix can be directly related to measurable final-state lepton distributions.
As shown in Fig.~\ref{fig:illustratonPlot}, in the c.m.~frame, photon conversion kinematics can be parametrized by the lepton azimuthal angle $\phi$ and the light-cone momentum fraction
\begin{equation}
z=\frac{E_\ell+p^z_\ell}{E_\gamma+p_\gamma^z}.
\end{equation}
In the collinear limit with small virtuality $s\gg q^2$, $z$ is approximately the energy fraction $E_\ell/E_\gamma$. 
In terms of these variables, the matrix $\Gamma_{\lambda\lambda^\prime}$ in Eq.~\eqref{eq:Gamma0} can be decomposed in the Pauli matrix basis as
\begin{align}\label{eq:Gamma}
    \frac{1}{16q^2}\Gamma_{\lambda\lambda^\prime} = & ~D(z,q^2) \delta_{\lambda\lambda^\prime} +\\ 
    &H(z,q^2)\Big[\cos(2\phi) (\sigma_x)_{\lambda\lambda^\prime}+\sin(2\phi)(\sigma_y)_{\lambda\lambda^\prime}\Big]. \nonumber
\end{align}
Here, we define the functions $D(z,q^2)$ and $H(z,q^2)$ for $\gamma^*\to \ell^+\ell^-$ in analogy to the fragmentation functions,
\begin{subequations}
\begin{align}
   D(z,q^2) &=\frac{1}{2}+ \frac{m_\ell^2}{q^2}-z(1-z), \\
   H(z,q^2) &= \frac{m_\ell^2}{q^2}-z(1-z), 
\end{align}
\end{subequations}
which, respectively, characterize the overall unpolarized conversion probability and the linear polarization dependence.
Since the conversion process is invariant under charge conjugation, Eq.~\eqref{eq:Gamma} is invariant under $z\to(1-z)$ and $\phi\to\phi+\pi$.
Consequently, $\phi$ and $z$ can be defined without specifying the charge of $\ell^+$ or $\ell^-$.

\begin{figure}
    \centering
    \hspace*{-0.6cm}
    \includegraphics[width=0.45\textwidth]{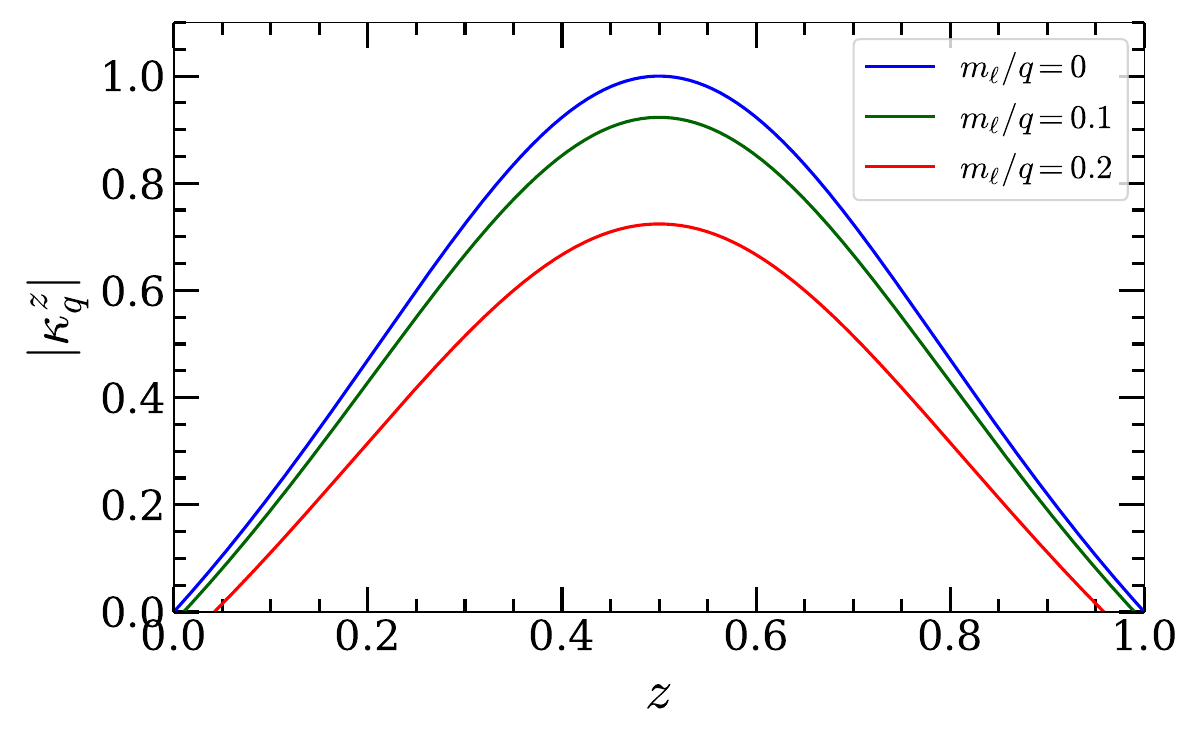}
    \caption{Spin analyzing power as a function of splitting kinematics.}
    \label{fig:HzqDzq}
\end{figure}

To parametrize how sensitive the final state distribution depends on linear polarization, it is convenient to introduce the spin-analyzing power
\begin{equation}
\kappa^z_q =  {H(z,q^2)\over D(z,q^2)}.   
\end{equation}
When $z=1/2$, the leptons have the largest transverse momentum to probe the azimuthal distribution and the spin analyzing power reaches the maximum, as shown in Fig.~\ref{fig:HzqDzq}. It decreases with $m_\ell/q$ because the lepton pair tends to be co-moving without angle distributions with increasing masses.

The structure of Eq.~\eqref{eq:Gamma} also reveals that no $\sigma_z$ term appears in the splitting density matrix, which is a result of parity conservation of QED processes. Unlike quantum tomography utilizing parity-violating decays that can reconstruct the full spin density matrix~\cite{Afik:2020onf}, the degree of circular polarization of the photon cannot be accessed through the splitting process. Nevertheless, the full information on the linear polarization can be reconstructed and it is sufficient to construct observables sensitive to quantum correlations in two-photon systems. 

With Eq.~\eqref{eq:Gamma} and the parameterization of the photon pair spin density matrix, the differential distribution in Eq.~\eqref{eq:M2fromRho} can be rewritten as
\begin{align}\label{eq:fullDistribution}
    \frac{\mathrm{d} \sigma}{ \mathrm{d}\Omega_1  \mathrm{d}\Omega_2} 
    &\propto \frac{D(z_1,q_1^2)D(z_2,q_2^2)}{q_1^2q_2^2} \sigma_{\rm hard} 
 \Bigg[1+ \alpha^{z_1,z_2}_{q_1,q_2}\times \nonumber\\
     \Big(&\mathcal{B}_- \cos2(\phi_1+\phi_2) + \mathcal{B}_+ \cos2(\phi_1-\phi_2)\Big)\Bigg].
\end{align}
Here, $\mathrm{d}\Omega_i =\mathrm{d}z_i \mathrm{d}q_i^2 \mathrm{d}\phi_i $ is the splitting phase space element, $\sigma_{\rm hard}$ is the cross section of the production of the real photon pair, and $\alpha^{z_1,z_2}_{q_1,q_2}= \kappa^{z_1}_{q_1}\kappa^{z_2}_{q_2}/2$ is the event-dependent spin analysis power.

The Bell variable can be directly reconstructed from the azimuthal distribution of the lepton pairs as 
\begin{equation}
\label{eq:Bellvariable}
    \mathcal{B}_\mp =2  \frac{\langle\cos2(\phi_1\pm\phi_2)\rangle}{\langle\alpha^{z_1,z_2}_{q_1,q_2}\rangle},
\end{equation}
where $\langle \mathcal{O} \rangle$ denotes the event average of the quantity $\mathcal{O}$. With $N$ reconstructed events ($N\gg 1$), we estimate the statistical uncertainty $\delta \mathcal{B}$ and the significance as 
\begin{equation}
\label{eq:delta}
\delta \mathcal{B} =  {\sqrt{2/N} \over \langle\alpha^{z_1,z_2}_{q_1,q_2}\rangle},\quad  \#\sigma =  {(\mathcal{B}-\sqrt{2}) \over \delta \mathcal{B}} .
\end{equation}
Alternatively, the inverse of the statistical significance $1/\#\sigma$ serves as a measure of the precision of the measurement.

\vskip 0.8mm
\noindent 
\emph{Simulation and sensitivity ---}
We carry out a detailed study for the process $e^+e^-\to \gamma^*\gamma^*\to \ell_1^+\ell_1^- \ell_2^+\ell_2^- $ in the Belle experiments \cite{Belle-II:2010dht} with $\sqrt{s}=10.58~{\rm GeV}$. 
We perform a Monte-Carlo simulation using \textsc{MadGraph}~\cite{Alwall:2011uj}.
Signal events are selected to contain exactly four charged leptons with zero net charge. To suppress background, all final-state leptons are required to satisfy $E_\ell>0.1~{\rm GeV}$ and $|\cos\theta_\ell|<0.9$ in the c.m.~frame, which are within the acceptance of the Belle detector~\cite{Belle-II:2010dht}.
To enrich the event selection for near-on-shell photon pair conversion, the four-lepton event is selected by requiring $m_{ee}<0.5~{\rm GeV}$ for electron pairs and $m_{\mu\mu}<1~{\rm GeV}$ for muon pairs.
These requirements suppress non-factorizable contributions and ensure the validity of the qubit approximation. 
To reconstruct the azimuthal asymmetry, we additionally require a minimal angular separation between the leptons of each reconstructed virtual photon. Taking advantage of the superb angular resolution,
\footnote{The angle resolution of electron tracks at Belle and BaBar detectors can reach sub mrad level~\cite{Belle:2000cnh,BaBar:2001yhh}.} 
we adopt a low threshold for the separation between the two leptons $\Delta R = \sqrt{\Delta\phi^2+\Delta\eta^2}>0.001$. After event selection, the signal cross sections for the $2e^+2e^-$ ($ee$), $e^+e^-\mu^+\mu^-$ ($e\mu$) and $2\mu^+2\mu^-$ ($\mu\mu$) channels are $0.15~{\rm pb}$, $0.076~{\rm pb}$ and $9.8\times 10^{-3}~{\rm pb}$, respectively. The residual background contributions remain at the level of approximately $7\%$, $4\%$ and $1\%$ for $ee$, $e\mu$ and $\mu\mu$ channels, respectively. 
For the same-flavor $ee$ and $\mu\mu$ channels, possible combinatorial ambiguities in pairing leptons into virtual photon candidates are below 0.001\% and 1\% levels, respectively. 
The background and mis-pairing in total modify the reconstructed Bell variable by less than 5\%, which do not qualitatively affect the sensitivity to Bell inequality violation and are therefore neglected in the following analysis.

\begin{figure}
    \centering
    \includegraphics[width=0.9\linewidth]{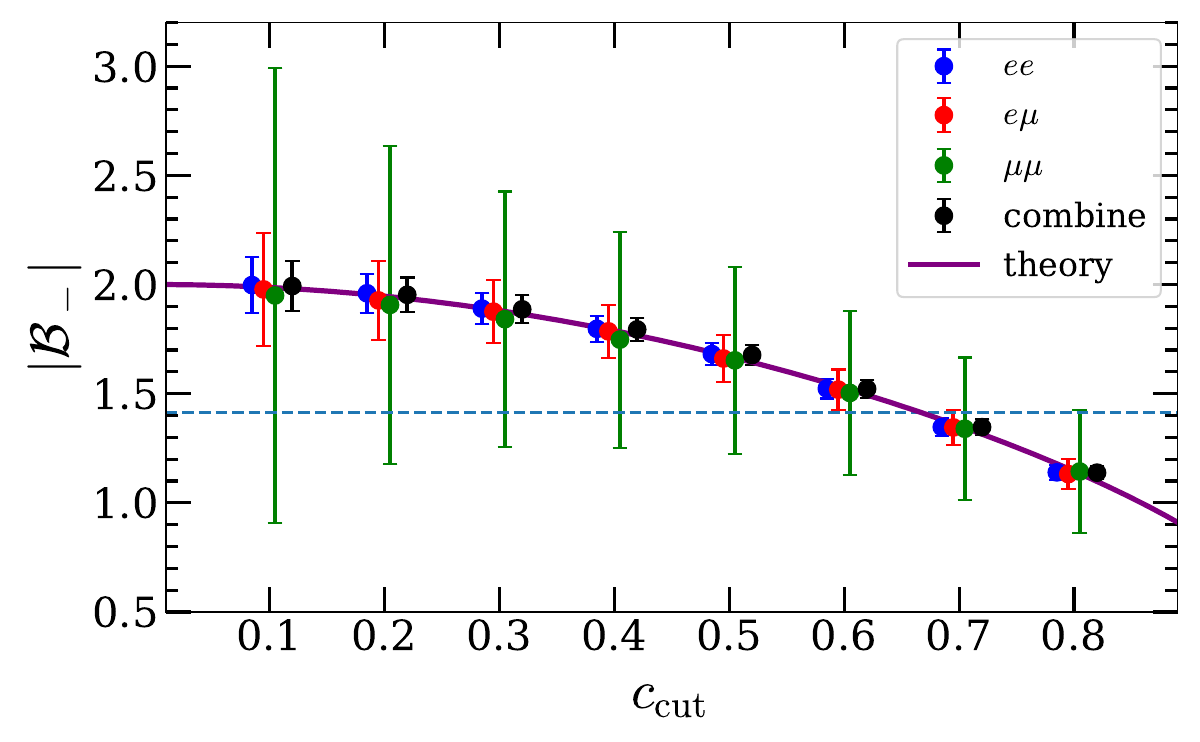}
    \caption{Reconstructed Bell variable $|\mathcal{B}_-|$ at Belle from photon conversion processes for different angular selections $|\cos\theta|<c_{\rm cut}$ and the requirement $\Delta R>0.001$. The statistical uncertainties are estimated for an integrated luminosity of $\mathcal{L}=1~\mathrm{ab}^{-1}$. The solid curve is from the prediction of the kinematic method in Eq.~(\ref{eq:BellSM}).}
    \label{fig:thetaCut}
\end{figure}

Using the selected event sample, the Bell variable in Eq.~\eqref{eq:Bellvariable} and its associated uncertainty are reconstructed from the simulated events using the analytical expression for the average of $\alpha_{q_1,q_2}^{z_1,z_2}$. Figure~\ref{fig:thetaCut} shows the reconstructed values of $\mathcal{B}_-$ in $ee$ (blue), $e\mu$ (red) and $\mu\mu$ (green) channels for different selection cuts $|\cos\theta|<c_{\rm cut}$ of the photon angle $\theta$, with the statistical uncertainties scaled to an integrated luminosity of $1~{\rm ab}^{-1}$.

For real photon pair production, quantum tomography can also be achieved in the kinematic approach directly from the scattering angle $\theta$~\cite{Cheng:2024rxi}, with the Bell variable given by 
\begin{equation}\label{eq:BellSM}
    \mathcal{B}_{-}=\frac{2\sin^2\theta}{1+\cos^2\theta},\qquad \mathcal{B}_{+}=0,
\end{equation}
where $\mathcal{B_-}$ reaches its maximal value 2 at $\cos\theta =0$.
For comparison, we also show the prediction in the kinematic approach with the purple line in Fig.~\ref{fig:thetaCut}.
We find the reconstructed central values agree with the prediction by the kinematic method, and the longitudinal mode contributions for low $q^2$ are at $\sim1\%$ level or smaller, confirming the validity of the effective two-qubit description of the virtual photon pair. 

As predicted by Eq.~\eqref{eq:BellSM}, $\mathcal{B}_-$ is enhanced in the central scattering region. We consider a requirement $|\cos\theta|<0.3$, which further reduces the background contributions by approximately 30\%. 
With this selection, the significance of Bell inequality violation are $6.6\sigma$, $3.2\sigma$ and $0.7\sigma$ for the $ee$, $e\mu$ and $\mu\mu$ channels, respectively, at an integrated luminosity of $1~{\rm ab}^{-1}$ comparable to the Belle data sample~\cite{Belle:2012iwr}. 
The $\gamma^*\to e^+e^-$ channel is more sensitive than $\gamma^*\to \mu^+\mu^-$ because of a larger spin analyzing power, as shown in Fig.~\ref{fig:HzqDzq}. 
While detector effects may also reduce these significances, the much larger data set expected at the Belle II experiment would provide a robust environment for an experimental realization.

To further improve the statistical sensitivity, we combine the three channels using inverse-variance weighting,
\begin{equation}
\mathcal{B}_-=\frac{\sum_i  \omega_i\mathcal{B}_{-}^i}{\sum_i \omega_i},
\label{eq:combine}
\end{equation}
where  $i=ee,~e\mu,~\mu\mu$ and the weights are defined as $\omega_i=1/(\delta \mathcal{B}_{-}^i)^2$.  The variance of the combined Bell variable is given by ${\rm Var}(\mathcal{B}_-)=1/\sum_i \omega_i$. 
The combined results are also shown as black points in Fig.~\ref{fig:thetaCut}. For the benchmark selection $|\cos\theta|<0.3$, a $7.4\sigma$ violation of the Bell inequality could be established at $\mathcal{L}=1~{\rm ab}^{-1}$.

A similar analysis can also be performed at BEPC \cite{BESIII20fb} at $\sqrt{s}=3.773~{\rm GeV}$ using the $ee$ channel, since the two-qubit approximation breaks down for the muon final state. The same event selection as in the Belle analysis is adopted, except for a tighter requirement $m_{ee}<0.2~{\rm GeV}$ to further suppress backgrounds. Using the currently accumulated luminosity of $20~{\rm fb}^{-1}$~\cite{BESIII20fb}, the expected significance reaches $2.4\sigma$. The sensitivity can be further improved with future luminosity upgrades, reaching the $5\sigma$ level at $\mathcal{L}=88~{\rm fb}^{-1}$.

\vskip 0.8mm
\noindent 
\emph{Other quantum information variables---}
Although the photon conversion process does not directly provide access to the $z$ components in the spin density matrix, the spin of two photons is guaranteed to be 100\% negatively correlated along the $z$ axis ($C_{zz}=1$) for on-shell photon states, and the spin correlation matrix is diagonal in the helicity basis.
This structure allows any quantum information observable to be investigated with $C_{xx/yy}$ reconstructed from Eq.~\eqref{eq:Bellvariable}. Here, we consider quantum discord and quantum magic as representative examples, while other observables can be evaluated in a similar way.

\begin{figure}
    \centering
    \includegraphics[width=0.49\linewidth]{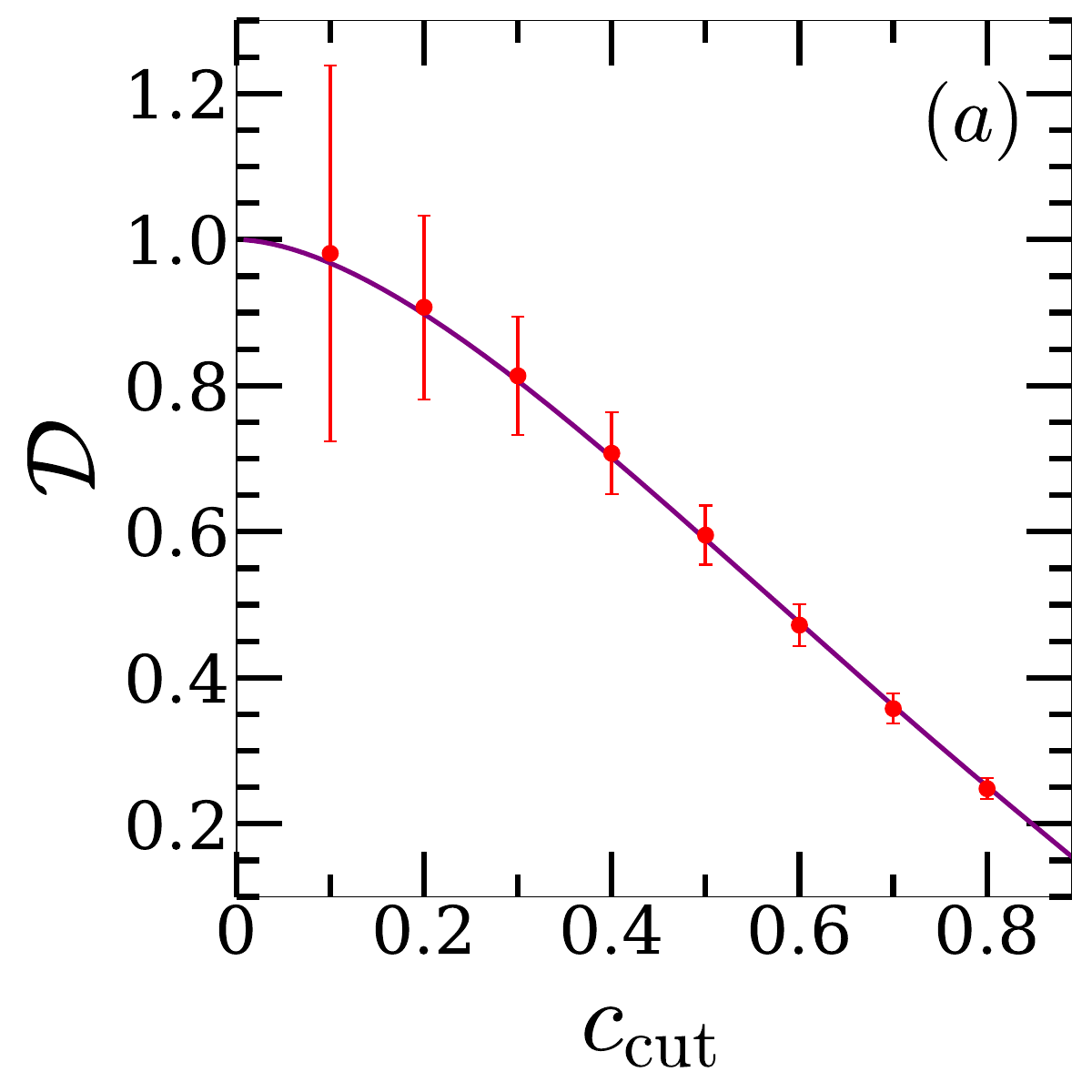}
    \includegraphics[width=0.49\linewidth]{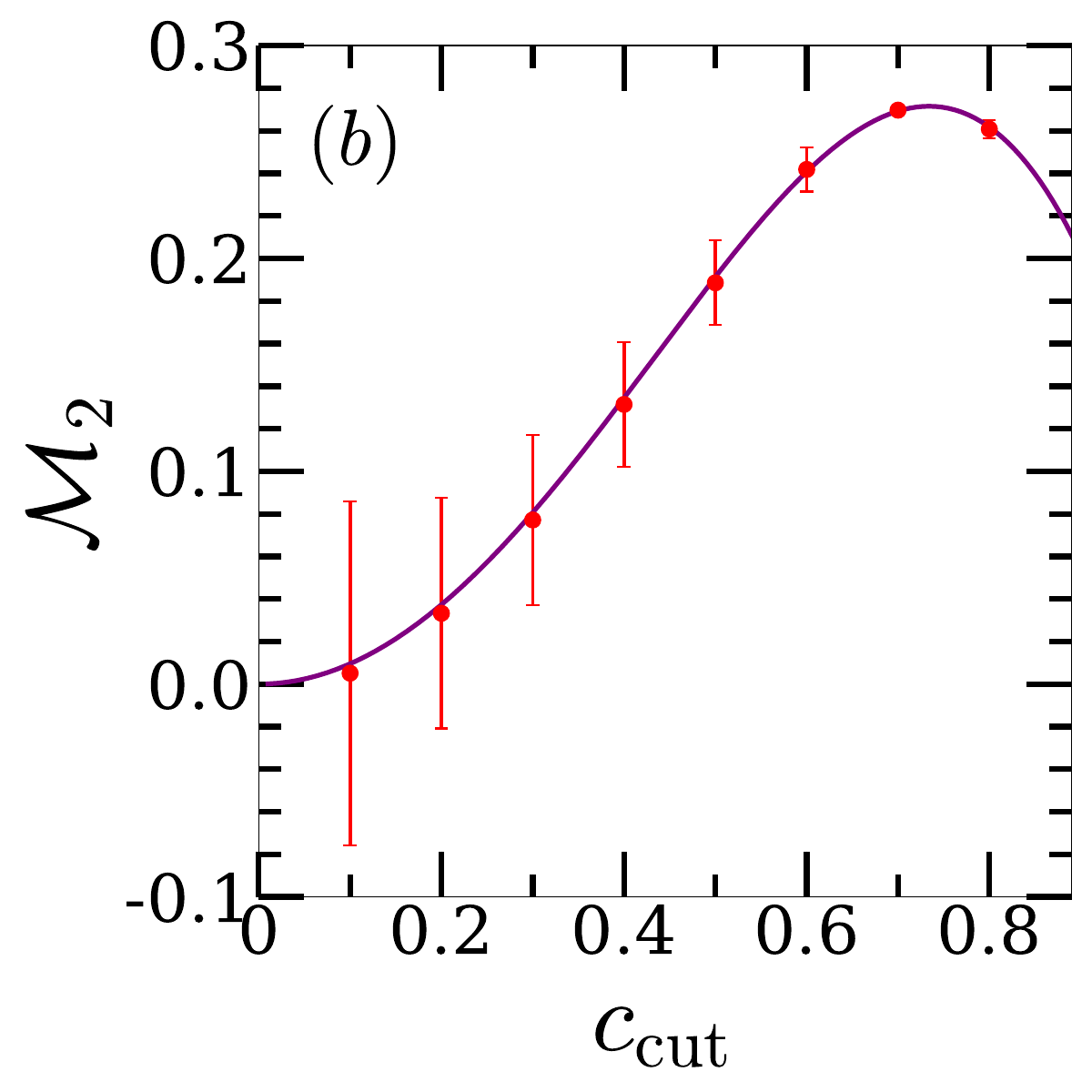}
    \caption{Reconstructed quantum discord $\mathcal{D}$ (a) and magic $\mathcal{M}_2$ (b), in comparison with prediction from the kinematic approach of real photon pair production.}
    \label{fig:DiscordM2}
\end{figure}

Quantum discord characterizes non-classical correlations more basic than entanglement.
Quantum discord is defined as the difference between two classically identical expressions for mutual information~\cite{Ollivier:2001fdq}, thus quantifying the non-classical correlation. For a two-qubit system, the discord reads~\cite{Luo:2008ecu}
\begin{equation}
\mathcal{D}  \equiv S(\rho_{\mathcal{B}})-S\left(\rho_{\mathcal{AB}}\right)  +\min_{\{\Pi_k\}}p_k S(\rho_{\mathcal{A}|k}),
\end{equation}
where $S(\rho)=-\operatorname{Tr}\left(\rho \log _2 \rho\right)$ is the von Neumann entropy, $\rho_{\mathcal{A/B}}=\tr_{\mathcal{B/A}}\left(\rho_{\mathcal{AB}}\right)$ is the reduced density matrix for the first/second qubit, and $\rho_{\mathcal{A}| k}$ represents the post-measurement quantum states for the qubit $\mathcal{A}$ given a projective measurement $\Pi_k$ on the qubit $\mathcal{B}$ with outcomes $k$ and projection probability $p_k$. The quantum discord can be calculated analytically when the spin correlation matrix is diagonal~\cite{Afik:2022dgh,Han:2024ugl,Wu:2025dds,Afik:2026pxv,Zhang:2026wvn}.
The results for different angular cuts, combining the $ee$, $e\mu$, and $\mu\mu$ channels at an integrated luminosity of $1~{\rm ab}^{-1}$ are shown in Fig.~\ref{fig:DiscordM2}(a). 
A nonzero discord, $0<\mathcal{D}\leq 1$, could be measured to a precision of $5.6\%$  for $c_{\text{cut}}=0.8$. We see again the consistency between the reconstructed results from the photon conversion and the solid line from the kinematic method.

Quantum magic describes the potential computational advantage of quantum states over classical states, and can also provide a complementary description of the state configuration. Any stabilizer state can be effectively simulated on a classical computer and has zero magic, whereas quantum states with higher magic provide a greater computational advantage over classical systems. Quantum magic can be described by second stabilizer R\'enyi entropy $\mathcal{M}_2$, which is evaluated as~\cite{White:2024nuc,Gargalionis:2025iqs,CMS:2025cim}
\begin{equation}
    \mathcal{M}_2 = - \log_2 \left( \frac{1+\sum_{i}(B_{i}^{(1)})^4 +\sum_{j}(B_{j}^{(2)})^4 +\sum_{i,j}C_{ij}^4}{1+\sum_{i}(B_{i}^{(1)})^2 +\sum_{j}(B_{j}^{(2)})^2 +\sum_{i,j}C_{ij}^2} \right).
\end{equation}
As shown in Fig.~\ref{fig:DiscordM2}(b), combining the $ee$, $e\mu$, and $\mu\mu$ channels at the same integrated luminosity, the quantum magic decreases with a smaller angular cut of $|\cos\theta|$, indicating that the state approaches a stabilizer Bell state in the central scattering region. The non-stabilizer nature could be measured to a precision of $1.6\%$ for $c_{\text{cut}}=0.8$. Again, the consistency between the solid line from the kinematic method and the reconstructed results shows the validity of the two-qubit framework.

\vskip 0.8mm
\noindent 
\emph{Conclusion---}
We proposed a framework to probe quantum correlations of energetic photon pairs via their conversion into dilepton pairs in the small-virtuality regime. In this limit, the conversion process factorizes from the hard production dynamics, enabling the photon polarization state to be treated as a qubit state and inferred from final-state angular correlations. 
We demonstrated the feasibility of this approach using the existing Belle data sample. Backgrounds from nonfactorizable effects are found to be negligible after event selections, while the effective two-qubit description remains valid at the percent level. The $ee$ splitting channel provides the most sensitive probe of quantum information due to its larger event yield and stronger spin-analyzing power. Combining the $ee$, $e\mu$, and $\mu\mu$ channels, a $7.4\sigma$ observation of Bell inequality violation could be achieved at $\mathcal{L}=1~{\rm ab}^{-1}$, while a positive quantum discord could be observed with a precision of 5.6\%. Quantum magic analysis indicates that the state approaches a stabilizer Bell state in the central region, and the departure from the stabilizer regime could be measured to a precision of 1.6\%. We found full consistency between the kinematic method and the reconstructed results from the photon conversion in the qubit approximation for spin-1 virtual photons. Our results establish photon conversion as a practical handle for accessing quantum properties of photons at colliders and identify photon pairs as a promising system for studying quantum entanglement in high-energy physics. The framework could also be extended to processes such as $h \to \gamma\gamma$ at the LHC as well as the other spin-1 states in an appropriate qubit limit. 

\vspace{3mm}
We thank C. Z. Yuan and M. Q. Ruan for helpful discussions. KC and TH were supported in part by the US Department of Energy under grant N.~DE-SC0007914, and in part by Pitt PACC. GL and BY were partly supported by the National Natural Science Foundation of China under Grant No.~12422506, and CAS under Grant No.~E429A6M1. The authors gratefully acknowledge the valuable discussions and insights provided by the members of the Collaboration on Precision Tests and New Physics (CPTNP).

\bibliographystyle{apsrev4-1}
\bibliography{ref}

@article{Clauser:1969ny,
    author = "Clauser, John F. and Horne, Michael A. and Shimony, Abner and Holt, Richard A.",
    title = "{Proposed experiment to test local hidden variable theories}",
    doi = "10.1103/PhysRevLett.23.880",
    journal = "Phys. Rev. Lett.",
    volume = "23",
    pages = "880--884",
    year = "1969"
}

@article{Weihs_1998,
   title={Violation of Bell’s Inequality under Strict Einstein Locality Conditions},
   volume={81},
   ISSN={1079-7114},
   url={http://dx.doi.org/10.1103/PhysRevLett.81.5039},
   DOI={10.1103/physrevlett.81.5039},
   number={23},
   journal={Physical Review Letters},
   publisher={American Physical Society (APS)},
   author={Weihs, Gregor and Jennewein, Thomas and Simon, Christoph and Weinfurter, Harald and Zeilinger, Anton},
   year={1998},
   month=Dec, pages={5039–5043} }

@article{kasday1975angular,
  title={Angular correlation of compton-scattered annihilation photons and hidden variables},
  author={Kasday, Leonard R and Ullman, Jack D and Wu, Chien-Shiung},
  doi="https://doi.org/10.1007/BF02724742",
  journal={Il Nuovo Cimento B (1971-1996)},
  volume={25},
  number={2},
  pages={633--661},
  year={1975},
  publisher={Springer}
}

@article{Wu:1950zz,
    author = "Wu, C. S. and Shaknov, I.",
    title = "{The Angular Correlation of Scattered Annihilation Radiation}",
    doi = "10.1103/PhysRev.77.136",
    journal = "Phys. Rev.",
    volume = "77",
    pages = "136--136",
    year = "1950"
}

@article{Freedman:1972zza,
    author = "Freedman, Stuart J. and Clauser, John F.",
    title = "{Experimental Test of Local Hidden-Variable Theories}",
    doi = "10.1103/PhysRevLett.28.938",
    journal = "Phys. Rev. Lett.",
    volume = "28",
    pages = "938--941",
    year = "1972"
}

@article{Aspect:1982fx,
    author = "Aspect, Alain and Dalibard, Jean and Roger, Gerard",
    title = "{Experimental test of Bell's inequalities using time varying analyzers}",
    doi = "10.1103/PhysRevLett.49.1804",
    journal = "Phys. Rev. Lett.",
    volume = "49",
    pages = "1804--1807",
    year = "1982"
}

@article{Verstraete:2001skx,
    author = "Verstraete, Frank and Wolf, Michael M.",
    title = "{Entanglement versus Bell Violations and Their Behavior under Local Filtering Operations}",
    eprint = "quant-ph/0112012",
    archivePrefix = "arXiv",
    doi = "10.1103/PhysRevLett.89.170401",
    journal = "Phys. Rev. Lett.",
    volume = "89",
    number = "17",
    pages = "170401",
    year = "2002"
}

@article{Wootters:1997id,
    author = "Wootters, William K.",
    title = "{Entanglement of formation of an arbitrary state of two qubits}",
    eprint = "quant-ph/9709029",
    archivePrefix = "arXiv",
    doi = "10.1103/PhysRevLett.80.2245",
    journal = "Phys. Rev. Lett.",
    volume = "80",
    pages = "2245--2248",
    year = "1998"
}

@article{PhysicsPhysiqueFizika.1.195,
    author = "Bell, J. S.",
    title = "{On the Einstein-Podolsky-Rosen paradox}",
    reportNumber = "RX-1376",
    doi = "10.1103/PhysicsPhysiqueFizika.1.195",
    journal = "Physics Physique Fizika",
    volume = "1",
    pages = "195--200",
    year = "1964"
}

@article{Afik:2025ejh,
    author = "Afik, Yoav and others",
    title = "{Quantum information meets high-energy physics: input to the update of the European strategy for particle physics}",
    eprint = "2504.00086",
    archivePrefix = "arXiv",
    primaryClass = "hep-ph",
    doi = "10.1140/epjp/s13360-025-06752-9",
    journal = "Eur. Phys. J. Plus",
    volume = "140",
    number = "9",
    pages = "855",
    year = "2025"
}

@article{Afik:2020onf,
    author = "Afik, Yoav and de Nova, Juan Ram\'on Mu\~noz",
    title = "{Entanglement and quantum tomography with top quarks at the LHC}",
    eprint = "2003.02280",
    archivePrefix = "arXiv",
    primaryClass = "quant-ph",
    doi = "10.1140/epjp/s13360-021-01902-1",
    journal = "Eur. Phys. J. Plus",
    volume = "136",
    number = "9",
    pages = "907",
    year = "2021"
}

@article{Fabbrichesi:2021npl,
    author = "Fabbrichesi, M. and Floreanini, R. and Panizzo, G.",
    title = "{Testing Bell Inequalities at the LHC with Top-Quark Pairs}",
    eprint = "2102.11883",
    archivePrefix = "arXiv",
    primaryClass = "hep-ph",
    doi = "10.1103/PhysRevLett.127.161801",
    journal = "Phys. Rev. Lett.",
    volume = "127",
    number = "16",
    pages = "161801",
    year = "2021"
}

@article{Severi:2021cnj,
    author = "Severi, Claudio and Boschi, Cristian Degli Esposti and Maltoni, Fabio and Sioli, Maximiliano",
    title = "{Quantum tops at the LHC: from entanglement to Bell inequalities}",
    eprint = "2110.10112",
    archivePrefix = "arXiv",
    primaryClass = "hep-ph",
    doi = "10.1140/epjc/s10052-022-10245-9",
    journal = "Eur. Phys. J. C",
    volume = "82",
    number = "4",
    pages = "285",
    year = "2022"
}

@article{Afik:2022kwm,
    author = "Afik, Yoav and de Nova, Juan Ram\'on Mu\~noz",
    title = "{Quantum information with top quarks in QCD}",
    eprint = "2203.05582",
    archivePrefix = "arXiv",
    primaryClass = "quant-ph",
    doi = "10.22331/q-2022-09-29-820",
    journal = "Quantum",
    volume = "6",
    pages = "820",
    year = "2022"
}

@article{Aguilar_Saavedra_2022,
    author = "Aguilar-Saavedra, J. A. and Casas, J. A.",
    title = "{Improved tests of entanglement and Bell inequalities with LHC tops}",
    eprint = "2205.00542",
    archivePrefix = "arXiv",
    primaryClass = "hep-ph",
    reportNumber = "IFT-UAM/CSIC-22-45",
    doi = "10.1140/epjc/s10052-022-10630-4",
    journal = "Eur. Phys. J. C",
    volume = "82",
    number = "8",
    pages = "666",
    year = "2022"
}

@article{Dong_2024,
    author = "Dong, Zhongtian and Gon\c{c}alves, Dorival and Kong, Kyoungchul and Navarro, Alberto",
    title = "{Entanglement and Bell inequalities with boosted tt\textasciimacron{}}",
    eprint = "2305.07075",
    archivePrefix = "arXiv",
    primaryClass = "hep-ph",
    doi = "10.1103/PhysRevD.109.115023",
    journal = "Phys. Rev. D",
    volume = "109",
    number = "11",
    pages = "115023",
    year = "2024"
}

@article{Han:2023fci,
    author = "Han, Tao and Low, Matthew and Wu, Tong Arthur",
    title = "{Quantum entanglement and Bell inequality violation in semi-leptonic top decays}",
    eprint = "2310.17696",
    archivePrefix = "arXiv",
    primaryClass = "hep-ph",
    reportNumber = "PITT-PACC-2316",
    doi = "10.1007/JHEP07(2024)192",
    journal = "JHEP",
    volume = "07",
    pages = "192",
    year = "2024"
}

@article{Cheng:2023qmz,
    author = "Cheng, Kun and Han, Tao and Low, Matthew",
    title = "{Optimizing fictitious states for Bell inequality violation in bipartite qubit systems with applications to the tt{\textasciimacron} system}",
    eprint = "2311.09166",
    archivePrefix = "arXiv",
    primaryClass = "hep-ph",
    reportNumber = "PITT-PACC-2321",
    doi = "10.1103/PhysRevD.109.116005",
    journal = "Phys. Rev. D",
    volume = "109",
    number = "11",
    pages = "116005",
    year = "2024"
}

@article{Cheng:2024btk,
    author = "Cheng, Kun and Han, Tao and Low, Matthew",
    title = "{Optimizing entanglement and Bell inequality violation in top antitop events}",
    eprint = "2407.01672",
    archivePrefix = "arXiv",
    primaryClass = "hep-ph",
    reportNumber = "PITT-PACC-2401",
    doi = "10.1103/PhysRevD.111.033004",
    journal = "Phys. Rev. D",
    volume = "111",
    number = "3",
    pages = "033004",
    year = "2025"
}

@article{Cheng:2024rxi,
    author = "Cheng, Kun and Han, Tao and Low, Matthew",
    title = "{Quantum tomography at colliders: With or without decays}",
    eprint = "2410.08303",
    archivePrefix = "arXiv",
    primaryClass = "hep-ph",
    reportNumber = "PITT-PACC-2408",
    doi = "10.1016/j.physletb.2025.139675",
    journal = "Phys. Lett. B",
    volume = "868",
    pages = "139675",
    year = "2025"
}

@article{Cheng:2025zcf,
    author = "Cheng, Kun and Han, Tao and Low, Matthew and Wu, Tong Arthur",
    title = "{Quantum Tomography in Neutral Meson and Antimeson Systems}",
    eprint = "2507.12513",
    archivePrefix = "arXiv",
    primaryClass = "hep-ph",
    month = "7",
    year = "2025"
}

@article{Afik:2025grr,
    author = "Afik, Yoav and Kats, Yevgeny and de Nova, Juan Ram{\'o}n Mu{\~n}oz and Soffer, Abner and Uzan, David",
    title = "{Entanglement and Bell nonlocality with bottom-quark pairs at hadron colliders}",
    eprint = "2406.04402",
    archivePrefix = "arXiv",
    primaryClass = "hep-ph",
    doi = "10.1103/fhkc-kfhr",
    journal = "Phys. Rev. D",
    volume = "111",
    number = "11",
    pages = "L111902",
    year = "2025"
}

@article{Ai:2025wnt,
    author = "Ai, Tengyu and Bi, Qi and He, Yuxin and Liu, Jia and Wang, Xiao-Ping",
    title = "{Ultimate Quantum Precision Limit at Colliders: Conditions and Case Studies}",
    eprint = "2506.10673",
    archivePrefix = "arXiv",
    primaryClass = "hep-ph",
    reportNumber = "CPTNP-2025-019",
    doi = "10.1103/3m4t-pk9b",
    journal = "Phys. Rev. Lett.",
    volume = "135",
    number = "24",
    pages = "241804",
    year = "2025"
}

@article{CMSCollaboration_2024,
    author = "Hayrapetyan, Aram and others",
    collaboration = "CMS",
    title = "{Observation of quantum entanglement in top quark pair production in proton\textendash{}proton collisions at $\sqrt{s} = 13$ TeV}",
    eprint = "2406.03976",
    archivePrefix = "arXiv",
    primaryClass = "hep-ex",
    reportNumber = "CMS-TOP-23-001, CERN-EP-2024-137",
    doi = "10.1088/1361-6633/ad7e4d",
    journal = "Rept. Prog. Phys.",
    volume = "87",
    number = "11",
    pages = "117801",
    year = "2024"
}

@article{ATLAS:2023fsd,
    author = "Aad, Georges and others",
    collaboration = "ATLAS",
    title = "{Observation of quantum entanglement with top quarks at the ATLAS detector}",
    eprint = "2311.07288",
    archivePrefix = "arXiv",
    primaryClass = "hep-ex",
    reportNumber = "CERN-EP-2023-230",
    doi = "10.1038/s41586-024-07824-z",
    journal = "Nature",
    volume = "633",
    number = "8030",
    pages = "542--547",
    year = "2024"
}

@article{Altakach:2022ywa,
    author = "Altakach, Mohammad Mahdi and Lamba, Priyanka and Maltoni, Fabio and Mawatari, Kentarou and Sakurai, Kazuki",
    title = "{Quantum information and CP measurement in H{\textrightarrow}{\ensuremath{\tau}}+{\ensuremath{\tau}}- at future lepton colliders}",
    eprint = "2211.10513",
    archivePrefix = "arXiv",
    primaryClass = "hep-ph",
    doi = "10.1103/PhysRevD.107.093002",
    journal = "Phys. Rev. D",
    volume = "107",
    number = "9",
    pages = "093002",
    year = "2023"
}

@article{Ehataht:2023zzt,
    author = {Ehat{\"a}ht, Karl and Fabbrichesi, Marco and Marzola, Luca and Veelken, Christian},
    title = "{Probing entanglement and testing Bell inequality violation with e+e-{\textrightarrow}{\ensuremath{\tau}}+{\ensuremath{\tau}}- at Belle II}",
    eprint = "2311.17555",
    archivePrefix = "arXiv",
    primaryClass = "hep-ph",
    doi = "10.1103/PhysRevD.109.032005",
    journal = "Phys. Rev. D",
    volume = "109",
    number = "3",
    pages = "032005",
    year = "2024"
}

@article{Ma:2023yvd,
    author = "Ma, Kai and Li, Tong",
    title = "{Testing Bell inequality through ${\boldsymbol h{\bf\rightarrow}\boldsymbol\tau\boldsymbol\tau }$ at CEPC*}",
    eprint = "2309.08103",
    archivePrefix = "arXiv",
    primaryClass = "hep-ph",
    doi = "10.1088/1674-1137/ad62d8",
    journal = "Chin. Phys. C",
    volume = "48",
    number = "10",
    pages = "103105",
    year = "2024"
}

@article{Fabbrichesi:2024wcd,
    author = "Fabbrichesi, M. and Marzola, L.",
    title = "{Quantum tomography with {\ensuremath{\tau}} leptons at the FCC-ee: Entanglement, Bell inequality violation, sin{\ensuremath{\theta}}W, and anomalous couplings}",
    eprint = "2405.09201",
    archivePrefix = "arXiv",
    primaryClass = "hep-ph",
    doi = "10.1103/PhysRevD.110.076004",
    journal = "Phys. Rev. D",
    volume = "110",
    number = "7",
    pages = "076004",
    year = "2024"
}

@article{Han:2025ewp,
    author = "Han, Tao and Low, Matthew and Su, Youle",
    title = "{Entanglement and Bell Nonlocality in $\tau^+ \tau^-$ at the BEPC}",
    eprint = "2501.04801",
    archivePrefix = "arXiv",
    primaryClass = "hep-ph",
    reportNumber = "PITT-PACC-2412",
    month = "1",
    year = "2025"
}

@article{Zhang:2025mmm,
    author = "Zhang, Yulei and Zhou, Bai-Hong and Liu, Qi-Bin and Li, Shu and Hsu, Shih-Chieh and Han, Tao and Low, Matthew and Wu, Tong Arthur",
    title = "{Entanglement and Bell Nonlocality in $\tau^+ \tau^-$ at the LHC using Machine Learning for Neutrino Reconstruction}",
    eprint = "2504.01496",
    archivePrefix = "arXiv",
    primaryClass = "hep-ph",
    month = "4",
    year = "2025"
}

@article{Barr:2021zcp,
    author = "Barr, Alan J.",
    title = "{Testing Bell inequalities in Higgs boson decays}",
    eprint = "2106.01377",
    archivePrefix = "arXiv",
    primaryClass = "hep-ph",
    doi = "10.1016/j.physletb.2021.136866",
    journal = "Phys. Lett. B",
    volume = "825",
    pages = "136866",
    year = "2022"
}

@article{Barr:2022wyq,
    author = "Barr, Alan J. and Caban, Pawel and Rembieli\'nski, Jakub",
    title = "{Bell-type inequalities for systems of relativistic vector bosons}",
    eprint = "2204.11063",
    archivePrefix = "arXiv",
    primaryClass = "quant-ph",
    doi = "10.22331/q-2023-07-27-1070",
    journal = "Quantum",
    volume = "7",
    pages = "1070",
    year = "2023"
}

@article{Ashby-Pickering:2022umy,
    author = "Ashby-Pickering, Rachel and Barr, Alan J. and Wierzchucka, Agnieszka",
    title = "{Quantum state tomography, entanglement detection and Bell violation prospects in weak decays of massive particles}",
    eprint = "2209.13990",
    archivePrefix = "arXiv",
    primaryClass = "quant-ph",
    doi = "10.1007/JHEP05(2023)020",
    journal = "JHEP",
    volume = "05",
    pages = "020",
    year = "2023"
}

@article{Aguilar-Saavedra:2022wam,
    author = "Aguilar-Saavedra, J. A. and Bernal, A. and Casas, J. A. and Moreno, J. M.",
    title = "{Testing entanglement and Bell inequalities in H\textrightarrow{}ZZ}",
    eprint = "2209.13441",
    archivePrefix = "arXiv",
    primaryClass = "hep-ph",
    doi = "10.1103/PhysRevD.107.016012",
    journal = "Phys. Rev. D",
    volume = "107",
    number = "1",
    pages = "016012",
    year = "2023"
}

@article{Fabbrichesi:2023cev,
    author = "Fabbrichesi, Marco and Floreanini, Roberto and Gabrielli, Emidio and Marzola, Luca",
    title = "{Bell inequalities and quantum entanglement in weak gauge boson production at the LHC and future colliders}",
    eprint = "2302.00683",
    archivePrefix = "arXiv",
    primaryClass = "hep-ph",
    doi = "10.1140/epjc/s10052-023-11935-8",
    journal = "Eur. Phys. J. C",
    volume = "83",
    number = "9",
    pages = "823",
    year = "2023"
}

@article{Fabbri:2023ncz,
    author = "Fabbri, Federica and Howarth, James and Maurin, Theo",
    title = "{Isolating semi-leptonic $H\rightarrow WW^{*}$decays for Bell inequality tests}",
    eprint = "2307.13783",
    archivePrefix = "arXiv",
    primaryClass = "hep-ph",
    doi = "10.1140/epjc/s10052-023-12371-4",
    journal = "Eur. Phys. J. C",
    volume = "84",
    number = "1",
    pages = "20",
    year = "2024"
}

@article{Bi:2023uop,
    author = "Bi, Qi and Cao, Qing-Hong and Cheng, Kun and Zhang, Hao",
    title = "{New observables for testing Bell inequalities in W boson pair production}",
    eprint = "2307.14895",
    archivePrefix = "arXiv",
    primaryClass = "hep-ph",
    doi = "10.1103/PhysRevD.109.036022",
    journal = "Phys. Rev. D",
    volume = "109",
    number = "3",
    pages = "036022",
    year = "2024"
}

@article{Bernal:2024xhm,
    author = "Bernal, Alexander and Caban, Pawe{\l} and Rembieli{\'n}ski, Jakub",
    title = "{Entanglement and Bell inequality violation in vector diboson systems produced in decays of spin-0 particles}",
    eprint = "2405.16525",
    archivePrefix = "arXiv",
    primaryClass = "hep-ph",
    doi = "10.1038/s41598-025-07747-3",
    journal = "Sci. Rep.",
    volume = "15",
    number = "1",
    pages = "23410",
    year = "2025"
}

@article{Grossi:2024jae,
    author = "Grossi, Michele and Pelliccioli, Giovanni and Vicini, Alessandro",
    title = "{From angular coefficients to quantum observables: a phenomenological appraisal in di-boson systems}",
    eprint = "2409.16731",
    archivePrefix = "arXiv",
    primaryClass = "hep-ph",
    reportNumber = "COMETA-2024-24, MPP-2024-183, TIF-UNIMI-2024-15",
    doi = "10.1007/JHEP12(2024)120",
    journal = "JHEP",
    volume = "12",
    pages = "120",
    year = "2024"
}

@article{Goncalves:2025qem,
    author = "Gon{\c{c}}alves, Dorival and Kaladharan, Ajay and Krauss, Frank and Navarro, Alberto",
    title = "{Quantum Entanglement is Quantum: ZZ Production at the LHC}",
    eprint = "2505.12125",
    archivePrefix = "arXiv",
    primaryClass = "hep-ph",
    month = "5",
    year = "2025"
}

@article{Goncalves:2025xer,
    author = "Gon{\c{c}}alves, Dorival and Kaladharan, Ajay and Navarro, Alberto",
    title = "{Higher-Order Corrections to Quantum Observables in $h\to WW^*$}",
    eprint = "2506.19951",
    archivePrefix = "arXiv",
    primaryClass = "hep-ph",
    month = "6",
    year = "2025"
}

@article{Goncalves:2026njf,
    author = "Gon{\c{c}}alves, Dorival and Kaladharan, Ajay and Navarro, Alberto",
    title = "{Quantum Tomography and Entanglement in Semi-Leptonic $h\to VV^*$ Decays at Higher Orders}",
    eprint = "2604.16218",
    archivePrefix = "arXiv",
    primaryClass = "hep-ph",
    month = "4",
    year = "2026"
}

@article{Morales:2023gow,
   author = "Morales, R. A.",
   title = "{Exploring Bell inequalities and quantum entanglement in
vector boson scattering}",
   eprint = "2306.17247",
   archivePrefix = "arXiv",
   primaryClass = "hep-ph",
   doi = "10.1140/epjp/s13360-023-04784-7",
   journal = "Eur. Phys. J. Plus",
   volume = "138",
   number = "12",
   pages = "1157",
   year = "2023"
}

@article{Morales:2024jhj,
   author = "Morales, R. A.",
   title = "{Tripartite entanglement and Bell non-locality in
loop-induced Higgs boson decays}",
   eprint = "2403.18023",
   archivePrefix = "arXiv",
   primaryClass = "hep-ph",
   doi = "10.1140/epjc/s10052-024-12921-4",
   journal = "Eur. Phys. J. C",
   volume = "84",
   number = "6",
   pages = "581",
   year = "2024"
}

@article{Aguilar-Saavedra:2022mpg,
    author = "Aguilar-Saavedra, J. A.",
    title = "{Laboratory-frame tests of quantum entanglement in H{\textrightarrow}WW}",
    eprint = "2209.14033",
    archivePrefix = "arXiv",
    primaryClass = "hep-ph",
    reportNumber = "IFT-UAM/CSIC-22-119",
    doi = "10.1103/PhysRevD.107.076016",
    journal = "Phys. Rev. D",
    volume = "107",
    number = "7",
    pages = "076016",
    year = "2023"
}

@article{Aguilar-Saavedra:2025byk,
    author = "Aguilar-Saavedra, J. A.",
    title = "{Quantum tomography beyond the leading order}",
    eprint = "2505.11870",
    archivePrefix = "arXiv",
    primaryClass = "hep-ph",
    reportNumber = "IFT-UAM/CSIC-25-49",
    doi = "10.1140/epjc/s10052-025-14710-z",
    journal = "Eur. Phys. J. C",
    volume = "85",
    number = "9",
    pages = "969",
    year = "2025"
}

@article{ATLAS:2026hye,
    author = "Aad, Georges and others",
    collaboration = "ATLAS",
    title = "{Measurements of $Z$-boson pair entanglement in decays of Higgs bosons at the ATLAS experiment}",
    eprint = "2603.26463",
    archivePrefix = "arXiv",
    primaryClass = "hep-ex",
    reportNumber = "CERN-EP-2026-061",
    month = "3",
    year = "2026"
}

@article{CMS:2025anw,
    collaboration = "CMS",
    title = "{Study of spin correlations in Higgs boson decays to four leptons at CMS}",
    reportNumber = "CMS-PAS-HIG-25-011",
    year = "2025"
}

@article{Cheng:2025cuv,
    author = "Cheng, Kun and Yan, Bin",
    title = "{Bell Inequality Violation of Light Quarks in Dihadron Pair Production at Lepton Colliders}",
    eprint = "2501.03321",
    archivePrefix = "arXiv",
    primaryClass = "hep-ph",
    reportNumber = "PITT-PACC-2414",
    doi = "10.1103/gmqz-v4cl",
    journal = "Phys. Rev. Lett.",
    volume = "135",
    number = "1",
    pages = "011902",
    year = "2025"
}

@article{Lin:2025eci,
    author = "Lin, Shi-Jia and Liu, Ming-Jun and Shao, Ding Yu and Wei, Shu-Yi",
    title = "{Spin correlations and Bell nonlocality in $\Lambda\bar\Lambda$ wpair production from $e^+e^-$ collisions with a thrust cut}",
    eprint = "2507.15387",
    archivePrefix = "arXiv",
    primaryClass = "hep-ph",
    month = "7",
    year = "2025"
}

@article{Pei:2025yvr,
    author = "Pei, Junle and Hao, Xiqing and Wang, Xiaochuan and Li, Tianjun",
    title = "{Observation of quantum entanglement in $\Lambda \bar{\Lambda}$ pair production via electron-positron annihilation}",
    eprint = "2505.09931",
    archivePrefix = "arXiv",
    primaryClass = "hep-ph",
    month = "5",
    year = "2025"
}

@article{Pei:2025ito,
    author = "Pei, Junle and Li, Tianjun and Wu, Lina and Hao, Xiqing and Wang, Xiaochuan",
    title = "{Quantum Entanglement without Spin-Analyzing Power Dependence at the Colliders}",
    eprint = "2510.08031",
    archivePrefix = "arXiv",
    primaryClass = "hep-ph",
    month = "10",
    year = "2025"
}

@article{Pei:2026wfu,
    author = "Pei, Junle and Wu, Lina and Li, Tianjun and Hao, Xiqing",
    title = "{Excluding Local Hidden Variables in $\Lambda\bar \Lambda$ Production: The Incompatibility with Angular-Momentum Conservation and CPT Invariance}",
    eprint = "2601.15747",
    archivePrefix = "arXiv",
    primaryClass = "hep-ph",
    month = "1",
    year = "2026"
}

@article{Chen:2013epa,
    author = "Chen, Shion and Nakaguchi, Y{\={u}}ki and Komamiya, Sachio",
    title = "{Testing Bell's Inequality using Charmonium Decays}",
    eprint = "1302.6438",
    archivePrefix = "arXiv",
    primaryClass = "hep-ph",
    doi = "10.1093/ptep/ptt032",
    journal = "PTEP",
    volume = "2013",
    number = "6",
    pages = "063A01",
    year = "2013"
}

@article{Du:2024sly,
    author = "Du, Yong and He, Xiao-Gang and Liu, Chia-Wei and Ma, Jian-Ping",
    title = "{Impact of parity violation on quantum entanglement and Bell nonlocality}",
    eprint = "2409.15418",
    archivePrefix = "arXiv",
    primaryClass = "hep-ph",
    month = "9",
    year = "2024"
}

@article{Cao:2025qua,
    author = "Cao, Qing-Hong and Li, Guanghui and Wen, Xin-Kai and Yan, Bin",
    title = "{Probing Quark Electromagnetic Properties via Entangled Quark Pairs in Fragmentation Hadrons at Lepton Colliders}",
    eprint = "2509.18276",
    archivePrefix = "arXiv",
    primaryClass = "hep-ph",
    reportNumber = "CPTNP-2025-035",
    month = "9",
    year = "2025"
}

@article{Qi:2025onf,
    author = "Qi, Wei and Guo, Zijing and Xiao, Bo-Wen",
    title = "{Studying maximal entanglement and Bell nonlocality at an electron-ion collider}",
    eprint = "2506.12889",
    archivePrefix = "arXiv",
    primaryClass = "hep-ph",
    doi = "10.1103/6ycn-x3yj",
    journal = "Phys. Rev. D",
    volume = "113",
    number = "5",
    pages = "054048",
    year = "2026"
}

@article{Cheng:2025zaw,
    author = "Cheng, Kun and Han, Tao and Trifinopoulos, Sokratis",
    title = "{Quantum Information at the Electron-Ion Collider}",
    eprint = "2510.23773",
    archivePrefix = "arXiv",
    primaryClass = "hep-ph",
    reportNumber = "PITT-PACC-2509, CERN-TH-2025-205, MIT-CTP/5943",
    month = "10",
    year = "2025"
}

@article{Fucilla:2025kit,
    author = "Fucilla, Michael and Hatta, Yoshitaka",
    title = "{Spin-spin entanglement in diffractive heavy-quark production}",
    eprint = "2509.05267",
    archivePrefix = "arXiv",
    primaryClass = "hep-ph",
    doi = "10.1103/gbk8-z3dd",
    journal = "Phys. Rev. D",
    volume = "113",
    number = "3",
    pages = "L031504",
    year = "2026"
}

@article{Hatta:2025obw,
    author = "Hatta, Yoshitaka and Schoenleber, Jakob",
    title = "{Probing quantum entanglement with generalized parton distributions at the Electron-Ion Collider}",
    eprint = "2511.04537",
    archivePrefix = "arXiv",
    primaryClass = "hep-ph",
    doi = "10.1103/qdb2-k2nh",
    journal = "Phys. Rev. D",
    volume = "113",
    number = "9",
    pages = "094016",
    year = "2026"
}

@article{Fucilla:2026mkg,
    author = "Fucilla, Michael and Hatta, Yoshitaka and Xiao, Bo-Wen",
    title = "{Quantum entanglement in electron-nucleus collisions: Role of the linearly polarized gluon distribution}",
    eprint = "2604.11697",
    archivePrefix = "arXiv",
    primaryClass = "hep-ph",
    month = "4",
    year = "2026"
}

@article{Guo:2026yhz,
    author = "Guo, Yu-Chen and Han, Tao and Low, Matthew and Su, Youle",
    title = "{Quantum Tomography of Fermion Pairs in $e^+e^-$ Collisions: Longitudinal Beam Polarization Effects}",
    eprint = "2602.02719",
    archivePrefix = "arXiv",
    primaryClass = "hep-ph",
    month = "2",
    year = "2026"
}

@article{Fang:2026ddi,
    author = "Fang, Yi-Jing and Bhoonah, Amit and Cheng, Kun and Han, Tao and Liu, Yandong and Zhang, Hao",
    title = "{Spin Correlation and Quantum Entanglement of Fermion Pairs in Transversely Polarized $e^-e^+$ Collisions}",
    eprint = "2604.11887",
    archivePrefix = "arXiv",
    primaryClass = "hep-ph",
    reportNumber = "PITT-PACC-2604",
    month = "4",
    year = "2026"
}

@article{Alwall:2011uj,
    author = "Alwall, Johan and Herquet, Michel and Maltoni, Fabio and Mattelaer, Olivier and Stelzer, Tim",
    title = "{MadGraph 5 : Going Beyond}",
    eprint = "1106.0522",
    archivePrefix = "arXiv",
    primaryClass = "hep-ph",
    reportNumber = "FERMILAB-PUB-11-448-T",
    doi = "10.1007/JHEP06(2011)128",
    journal = "JHEP",
    volume = "06",
    pages = "128",
    year = "2011"
}

@article{Ollivier:2001fdq,
    author = "Ollivier, Harold and Zurek, Wojciech H.",
    title = "{Introducing Quantum Discord}",
    eprint = "quant-ph/0105072",
    archivePrefix = "arXiv",
    reportNumber = "LA-UR-01-2845, LA-UR 01-2845",
    doi = "10.1103/PhysRevLett.88.017901",
    journal = "Phys. Rev. Lett.",
    volume = "88",
    number = "1",
    pages = "017901",
    year = "2001"
}

@article{Afik:2022dgh,
    author = "Afik, Yoav and de Nova, Juan Ram{\'o}n Mu{\~n}oz",
    title = "{Quantum Discord and Steering in Top Quarks at the LHC}",
    eprint = "2209.03969",
    archivePrefix = "arXiv",
    primaryClass = "quant-ph",
    doi = "10.1103/PhysRevLett.130.221801",
    journal = "Phys. Rev. Lett.",
    volume = "130",
    number = "22",
    pages = "221801",
    year = "2023"
}

@article{Luo:2008ecu,
    author = "Luo, Shunlong",
    title = "{Quantum discord for two-qubit systems}",
    doi = "10.1103/PhysRevA.77.042303",
    journal = "Phys. Rev. A",
    volume = "77",
    number = "4",
    pages = "042303",
    year = "2008"
}

@article{Han:2024ugl,
    author = "Han, Tao and Low, Matthew and McGinnis, Navin and Su, Shufang",
    title = "{Measuring quantum discord at the LHC}",
    eprint = "2412.21158",
    archivePrefix = "arXiv",
    primaryClass = "hep-ph",
    reportNumber = "PITT-PACC-2316",
    doi = "10.1007/JHEP05(2025)081",
    journal = "JHEP",
    volume = "05",
    pages = "081",
    year = "2025"
}

@article{Wu:2025dds,
    author = "Wu, Sihao and Qian, Chen and Wang, Qun and Yang, Yang-Guang",
    title = "{Quantum steering and discord in hyperon-antihyperon system in electron-positron annihilation}",
    eprint = "2509.14990",
    archivePrefix = "arXiv",
    primaryClass = "hep-ph",
    doi = "10.1103/x43x-3wgy",
    journal = "Phys. Rev. D",
    volume = "113",
    number = "5",
    pages = "056030",
    year = "2026"
}

@article{Afik:2026pxv,
    author = "Afik, Yoav and Demina, Regina and Herrera, Alan and Heinz Hindrichs, Otto and de Nova, Juan Ram{\'o}n Mu{\~n}oz and Ravina, Baptiste",
    title = "{Experimental characterization of the hierarchy of quantum correlations in top quark pairs}",
    eprint = "2602.15115",
    archivePrefix = "arXiv",
    primaryClass = "quant-ph",
    month = "2",
    year = "2026"
}

@article{Zhang:2026wvn,
    author = "Zhang, Hong-Wei and Cao, Xu and Feng, Tai-Fu",
    title = "{Controlling Quantum discord and steering in Electron-Positron Annihilation Using Polarized Beams}",
    eprint = "2605.19642",
    archivePrefix = "arXiv",
    primaryClass = "hep-ph",
    month = "5",
    year = "2026"
}

@article{White:2024nuc,
    author = "White, Chris D. and White, Martin J.",
    title = "{Magic states of top quarks}",
    eprint = "2406.07321",
    archivePrefix = "arXiv",
    primaryClass = "hep-ph",
    reportNumber = "ADP-24-10/T1249",
    doi = "10.1103/PhysRevD.110.116016",
    journal = "Phys. Rev. D",
    volume = "110",
    number = "11",
    pages = "116016",
    year = "2024"
}

@article{Gargalionis:2025iqs,
    author = "Gargalionis, John and Moynihan, Nathan and Trifinopoulos, Sokratis and Wallace, Ewan N. V. and White, Chris D. and White, Martin J.",
    title = "{Spin versus nonstabilizerness in gluon and graviton scattering}",
    eprint = "2508.14967",
    archivePrefix = "arXiv",
    primaryClass = "hep-th",
    reportNumber = "CERN-TH-2025-166, ADP-25-30/T1292",
    doi = "10.1103/tcb9-3vpr",
    journal = "Phys. Rev. D",
    volume = "113",
    number = "1",
    pages = "016007",
    year = "2026"
}

@article{CMS:2025cim,
    author = {{CMS collaboration}},
    title = {Observation of magic states of top quark pairs produced in proton-proton collisions at $\sqrt{s}=13~\mathrm{TeV}$},
    note = {{CMS-PAS-TOP-25-001}},
    year = {2025}
}

@article{Wang:2026nls,
    author = "Wang, Dianwei and Hao, Xiqing and Liu, Liwei and Wu, Lina and Li, Tianjun",
    title = "{Determining the Spin-Analyzing Powers via Invariants of the Spin Correlation Matrices and Probing the Bell Non-Locality at the Lepton Colliders}",
    eprint = "2605.13250",
    archivePrefix = "arXiv",
    primaryClass = "hep-ph",
    month = "5",
    year = "2026"
}

@article{Liu:2026dzv,
    author = "Liu, Yu-Xuan and Qi, Wei and He, Luo-Ting and Xiao, Bo-Wen",
    title = "{Polarization, Maximal Concurrence, and Pure States in High-Energy Collisions}",
    eprint = "2604.17756",
    archivePrefix = "arXiv",
    primaryClass = "hep-ph",
    month = "4",
    year = "2026"
}

@article{Zhang:2026nwm,
    author = "Zhang, Hong-Wei and Cao, Xu and Feng, Tai-Fu",
    title = "{Manipulating Bell nonlocality and entanglement in polarized electron-positron annihilation}",
    eprint = "2602.10389",
    archivePrefix = "arXiv",
    primaryClass = "hep-ph",
    month = "2",
    year = "2026"
}

@article{Belle-II:2010dht,
    author = "Abe, T. and others",
    collaboration = "Belle-II",
    title = "{Belle II Technical Design Report}",
    eprint = "1011.0352",
    archivePrefix = "arXiv",
    primaryClass = "physics.ins-det",
    reportNumber = "KEK-REPORT-2010-1",
    month = "11",
    year = "2010"
}

@misc{BESIII20fb,
  author = {{BESIII Collaboration}},
  title = {{BESIII} Data Taking at the $\psi(3770)$ Energy Region},
  year = {2024},
  note = {\url{https://english.ihep.cas.cn/nw/han/y24/202403/t20240305_657949.html}}
}

@article{Belle:2012iwr,
    author = "Brodzicka, Jolanta and others",
    collaboration = "Belle",
    title = "{Physics Achievements from the Belle Experiment}",
    eprint = "1212.5342",
    archivePrefix = "arXiv",
    primaryClass = "hep-ex",
    reportNumber = "KEK-REPORT-2012-5",
    doi = "10.1093/ptep/pts072",
    journal = "PTEP",
    volume = "2012",
    pages = "04D001",
    year = "2012"
}

@article{Du:2026qsy,
    author = "Du, Yong and Geng, Chao-Qiang and He, Xiao-Gang and Liu, Chia-Wei and Liu, Sheng-Lin and Liu, Xin-Yi",
    title = "{Spin-flavor entanglement in $Λ_b \to ΛD$ and weak phase extraction}",
    eprint = "2605.09682",
    archivePrefix = "arXiv",
    primaryClass = "hep-ph",
    month = "5",
    year = "2026"
}

@article{Yang:2026uwu,
    author = "Yang, Beizhi and Zhang, Yu and Wang, Zeren Simon and Zhou, Xiaorong",
    title = "{Entanglement measures and Bell-type spin-correlation observables in tau-lepton pairs at the Super Tau-Charm Facility}",
    eprint = "2603.05846",
    archivePrefix = "arXiv",
    primaryClass = "hep-ph",
    month = "3",
    year = "2026"
}

@article{Belle:2000cnh,
    author = "Abashian, A. and others",
    collaboration = "Belle",
    title = "{The Belle Detector}",
    reportNumber = "KEK-PROGRESS-REPORT-2000-4",
    doi = "10.1016/S0168-9002(01)02013-7",
    journal = "Nucl. Instrum. Meth. A",
    volume = "479",
    pages = "117--232",
    year = "2002"
}

\end{document}